\documentclass{article}
\usepackage{geometry}
\newcommand{\noun}[1]{\textsc{#1}}

\begin{document}

\title{An Intuitive Paradigm For Quantum Mechanics\\\emph{Physics Essays} \textbf{5}
(2) 226-234 (1992)}

\author{A. F. Kracklauer\thanks{
Department of Physics; University of Houston; Houston
}}

\date{}

\maketitle
\begin{abstract}
Use is made of a relativistic kinematic modulation effect to compliment imagery
from Stochastic Electrodynamics to provide intuitive paradigms for Quantum Mechanics.
Based on these paradigms, resolutions for epistemological problems vexing conventional
interpretations of Quantum Mechanics are proposed and discussed.
\end{abstract}

\section{Introduction}

For lack of an intuitive paradigm or physical model, Quantum Mechanics (QM)
remains an abstruse theory. Thus far all attempts to find underlying motivation
for its fundamental precepts which is somehow compatible with intuition, have
been found wanting. Moreover, some efforts to ``understand'' QM have led to
results indicating that it has conceptual features which are fundamentally irreconcilable
with macroscopic experience and classical physics. 

Bell's Theorem, for example, is understood nowadays to mean that certain aspects
of natural phenomena described by QM, are ``nonlocal.''\cite{1} That is, events
at a particular point with space-time coordinates \( x_{\mu } \), according
to the principles of QM, depend on the values of functions evaluated at points
outside the past light cone centered on \( x_{\mu } \). Because the classical
theories of fundamental physical interactions are all ``local'' in this sense,
this theorem appears to preclude the possibility that QM could be explained
at a deeper level with principles from classical physics, for example by insinuating
``hidden variables'' which are implicitly averaged out at the level of QM as
currently formulated. 

This implies in particular, that in spite of the probabilistic interpretation
given wave functions \( \psi (x) \), QM can not be just an elaboration on classical
Statistical Mechanics quantifying effects attributable to the accumulation of
finer-grained, fundamental interactions which are somehow governed by principles
from classical physics. Moreover, certain formalistic similarities to equations
from statistical physics notwithstanding, several arguments show that time evolution
in QM can not be described by a Markov process of the sort used in Stochastic
Mechanics. Markov processes are those for which time evolution has no long term
memory, as it were; i.e., trajectories can not be retrodicted. QM, to the contrary,
\emph{is} time-reversible---at least up to the point of measurement, an event
which, in any case, is not encompassed by the QM formalism for time evolution.
Thus, despite parallels with Statistical Mechanics, QM must encompass something
essentially different from irreversible (Markovian) Brownian motion, or similar
phenomena. 

The purpose of this article is to propose a new conceptual paradigm which, nevertheless,
does provide intuitive imagery from classical physics to motivate the basic
precepts of QM. From the viewpoint of this new model, the apparent nonlocal
aspects of QM are not objectionable because QM is reinterpreted, not as a fundamental
theory of elementary interactions, but as a theory of ``many-body effects''
attributable in a certain way to the accumulation of local, two-body, classical
electromagnetic interactions with the rest of the universe.

\section{Fundamental Concepts}

QM is formulated on two levels. The basic theory, or ``First Quantization,''
was conceived originally to describe the emission and absorption spectra of
atoms and molecules, and certain other phenomena, all having features suggesting
resonances in oscillatory systems or wave propagation. Thereafter, ``Second
Quantization'' was devised to describe refinements to the basic theory such
as the Lamb shift, the line width of atomic spectra, stimulated emission and
the anomalous magnetic moment of the electron, etc. Later it was seen that effects
described by Second Quantization often can be given imaginable, physical motivation
in terms of a \emph{Zitterbewegung}, or a fine-scale jitter similar to Brownian
motion and attributable to what often is called the ``quantized vacuum.'' This
observation, in turn, motivated various attempts to derive the fundamental equations
of QM or to describe quantum phenomena strictly on the basis of statistical
physics. Some of these attempts, under the rubric of Stochastic Electrodynamics
(SED), have successfully quantified certain phenomena usually described using
Second Quantization techniques. These phenomena include the Van Der Waals Forces,
the behavior of a charged nonrelativistic harmonic oscillator, and others, where,
not coincidently, the imagery of \emph{Zitterbewegung} appears germane.\cite{2}
However, the main corpus of QM, including atomic structure, spectral lines and
most strikingly, the wave-like diffraction of particle beams, continues to elude
all clarification in these terms. 

The basic, axiomatic premise underlying SED is that the universe is permeated
with random electromagnetic background radiation in thermodynamic equilibrium
with all charged particles. (For present purposes, a particle need not carry
a net charge, any multipole alone is sufficient.)\footnote{
On the other hand, particles with no charge structure, including no net charge,
no multipole moments, etc., are ``invisible'' to all forms of electromagnetic
radiation and are largely unmanipulatable by interaction with the material laboratory
world comprised of atoms and, as such, are inaccessible to direct experiment.
Their behavior remains, at best, a matter for indirect inference and is not
governed by the considerations presented herein. 
}The energy in this background radiation is envisioned to be a residue devolving
from past interactions or emissions from all charged particles in the universe.
For most derivations in classical physics this residue and its source are ignored;
for example, the derivations of the Thermodynamic formulae for an ideal gas
ignore radiation caused by acceleration incidental to collisions. Regardless
of the physical model ascribed to the generation of this background, however,
for the purposes of the formal theory or logical foundation of SED, its existence
is asserted as an axiomatic proposition.\cite{2}

The essential feature of background radiation is determined by the requirement
that its spectral energy density \( E(\omega ) \) be invariant under Lorentz
transformations in the sense that the total energy between two fixed numerical
values of \( \omega  \), \( a \) and \( b \), be identical in all inertial
frames; it i.e.,
\begin{equation}
\label{II.1}
\int ^{b}_{a}E(\omega )d^{3}k=\int ^{b}_{a}E'(\omega )\gamma (1-v/c)d^{3}k,\, \; \; [\gamma \equiv (1-v^{2}/c^{2})^{-1/2}]
\end{equation}
As Physics, this stipulation is tantamount to the requirement that there be
no distinguishable frames; and, it is based on the fact that, were it not true,
the background would engender certain anisotropisms that in fact are not observed.
Eq. (\ref{II.1}) is satisfied by a linear spectral energy density \( E(\omega )=constant\times \omega  \)
where the constant scale factor is determined empirically to be Planck's constant/\( 4\pi  \)---\( \hbar /2 \).
\cite{2}\cite{3}

Taking this Lorentz invariant background radiation into account leads directly
to the Planck blackbody spectrum and an understanding of ``photon statistics.''\cite{3}
This has been shown most simply by manipulation the following four equations
involving the mean energy density \( \overline{E_{i},} \) the mean square energy
density \( \overline{E^{2}_{i}} \), and the mean square deviation of the energy
density \( \overline{(\delta E_{i})^{2}} \) of any two mutually incoherent
radiation fields:
\begin{equation}
\label{II.2}
\overline{E_{sum}}=\overline{E_{1}}+\overline{E_{2}},
\end{equation}

\begin{equation}
\label{II.3}
\overline{E_{1}E_{2}}=\overline{E_{1}}\; \; \overline{E_{2}},
\end{equation}

\begin{equation}
\label{II.4}
\overline{(\delta E)^{2}}=\overline{(\delta E_{1})^{2}}+\overline{(\delta E_{2})^{2}},
\end{equation}
and

\begin{equation}
\label{II.5}
\overline{(\delta E_{i})^{2}}=\overline{E^{2}_{i}}-\overline{E_{i}}^{2}=\overline{E_{i}}^{2};\; \; \; i=1,2
\end{equation}
to obtain
\begin{equation}
\label{II.6}
\overline{(\delta E_{T})^{2}}=\overline{E^{2}_{T}}+2\, \overline{E_{T}}\; \; \overline{E_{B}},
\end{equation}
where, in the case at hand, \( \overline{E_{B}} \) is the mean energy density
solely of the background radiation, and \( \overline{E_{T}} \) is the same
for a temperature dependant radiation field. This latter field coexists with
the background and is modified by it via the mutual interference terms which
are included in \( \overline{E_{T}.} \) Invoking the Fluctuation Theorem :
\begin{equation}
\label{II.7}
\frac{\partial \, \overline{E_{T}}}{\partial \mu }=\overline{(\delta E_{T})^{2}},\; \; \; [\mu \equiv -1/\kappa T]
\end{equation}
for the thermal field at temperature \( T \) where \( \kappa  \) is Boltzmann's
constant, yields the differential equation:
\begin{equation}
\label{II.8}
\frac{\partial \, \overline{E_{T}}}{\partial \mu }=\overline{E_{T}}^{2}+2\, \overline{E_{T}}\; \overline{E_{B}},
\end{equation}
whose solution, with \( \overline{E_{B}}=\hbar \omega /2, \) is the Planck
black body spectral energy distribution
\begin{equation}
\label{II.9}
\overline{E_{T}}=\frac{\hbar \omega }{e^{\hbar \omega /\kappa T}-1}.
\end{equation}
The Fluctuation theorem is applicable if the thermal field exchanges only energy
with the background---a stipulation met here \emph{ab initio}. 

Furthermore, Eq. (\ref{II.6}) written in the form:
\begin{equation}
\label{II.10}
\frac{\overline{(\delta E_{T})^{2}}}{\overline{E_{T}}^{2}}=1+2\frac{\overline{E_{B}}}{\overline{E_{T}}}
\end{equation}
elucidates the source of the dualistic nature of radiation.\cite{3} Were the
first term on the right to stand alone, Eq. (\ref{II.10}) would characterize
intensity fluctuations of a classical radiation field while the second term
alone, being proportional to \( 1/\overline{E_{T}} \), gives an equation characterizing
density fluctuations of a particle ensemble. Together they capture the essence
of ``photon statistics'' and thus, without assuming the existence of discreet
quanta, characterize the ``quantized electromagnetic field.''\cite{3} 

\emph{Zitterbewegung}, in this theory, is interpreted as a manifestation of
the interaction of particles with the random, Lorentz-invariant background. 

First Quantization, however, pertains to resonance and wave-like phenomena for
which physical intuition admits no suggestion of \emph{Zitterbewegung}. However,
following SED, all particles with charge structure are considered to be in thermodynamic
equilibrium with the background via interaction with signals at frequencies
characteristic of their structure. The intuitive paradigm proposed herein to
model First Quantization is based on the \emph{Ansatz} that, in frames in which
they are moving, particles will be deflected by diffraction patterns in the
background signals to which they are tuned. Background waves, being conventional
electromagnetic radiation, are diffracted by physical boundaries according to
the usual principles of optics.\footnote{
Results from experiments in Cavity Quantum Electrodynamics \cite{4} substantiate
the concept of the background as ordinary electromagnetic radiation. Were background
radiation an immutable ground state or the ``quantized vacuum'' as conventionally
characterized by QM, then, contrary to observation, it might be expected to
resemble the vacuum (\emph{viz.}, an absolute space-time void) by being uniformly
ubiquitous and by permeating the interstices of materials down to the scale
of atoms and even ``elementary particles'' (essentially) without alteration,
in particular at macroscopic cavity boundaries. 
} For example, a particle moving towards a slit would equilibrate with a signal
that is a standing wave in its own frame but which is a traveling wave in the
slit's frame where it diffracts at boundaries such as those of the slit. On
passage through the slit, the particle is subject to the lateral energy flux
attendant to the diffraction of the background signal to which it is tuned.
In other words, it is envisioned that a particle will tend to be jostled into
the energy nodes of the diffraction pattern of the ``standing wave'' to which
it is tuned in its own inertial frame but which is a translating wave in the
frame of the slit. This effect is similar to the way froth and debris tend to
track the nodes of standing waves in rivers or sand tends to settle on the nodes
of a vibrating membrane. An ensemble of similar particles in identical circumstances---e.g.,
a beam of particles impinging on a slit---upon accumulation at the detector
discloses the diffraction pattern of the composite wave comprised of components
with which the individual particles are in equilibrium. 

Consider, for example, a neutral particle or system consisting of a dipole of
opposite charges held apart by some internal structure modeled to first order
by a simple spring with resonant frequency \( \omega _{0} \). According to
the basic SED assumption of thermodynamic equilibrium with the background, the
rest energy of this system constituting the particle will equal the energy in
the background mode \( \omega _{0} \) which is also the resonant frequency
of the system at which it is exchanging energy with the background; that is

\begin{equation}
\label{II.11}
m_{0}c^{2}=\hbar \omega _{0},
\end{equation}
where a contribution of \( \hbar \omega _{0}/2 \) is made to the right side
by both polarization states of the background mode. (In the case of systems
with more complex internal structure, \( \omega _{0} \) stands for the sum
of the frequencies corresponding to the various possible interactions.)\footnote{
Also, \( m_{0} \) is to be the sum of the masses of the system constituents
``renormalized'' to account for relativistic mass increases which are due to
internal motion. For present purposes, however, the details of the internal
structure of the particle are immaterial. 
} In its rest frame, with respect to each independent spatial direction, a particle
will equilibrate with a standing wave having an antinode at its location, which,
if the particle is located at \( x=0 \), has an intensity proportional to the
expression \( 2\cos (k_{0}x)\sin (\omega _{0}). \)\footnote{
Tautologically, a charged particle does not interact with those signals that
interfere destructively so as to have a node at its spacial location. 
} When projected onto a coordinate frame translating at velocity \( v \) with
respect to the particle, that of a slit for example, this standing wave has
the form of the modulated translating wave and is proportional to 
\begin{equation}
\label{II.12}
2\cos (k_{0}\gamma (x-c\beta t))\sin (\omega _{0}\gamma (t-c^{-1}\beta x));\; \; \; [\beta \equiv v/c]
\end{equation}
This wave consists of a short wavelength ``carrier'' modulated at a wavelength
\( \lambda =(\gamma \beta k_{0})^{-1} \) inversely proportional to the relative
velocity of the particle with respect to the slit. The modulation on this wave
is a relativistic kinematic effect. It arises from the difference in the Lorentz
transformations of the oppositely translating components of a standing wave. 

The modulated wave, upon propagation through a slit, for example, is diffracted
according to Huygens' principle such that the modulation diffraction pattern
is imposed on the carrier's diffraction pattern.\cite{5} A particle bathed
in this diffracted wave will experience a gross energy flux with a spatial pattern
proportional to the square of the modulation intensity imposed on the fine-scale
background wave driving the \emph{Zitterbewegung}. In other words, according
to this interpretation, boundary conditions on waves in the background modify
the stochastic effects of \emph{Zitterbewegung} on the orbits of material particles.
The actual detailed motion of a particle, while it reflects the relatively large
scale effects of the modulation, is very complex and jitters in consort with
spatially modulated \emph{Zitterbewegung}. 

Now, a Lorentz transformation into the translating frame applied to both sides
of the statement of energy equilibrium, Eq. (\ref{II.11}), yields both 
\begin{equation}
\label{II.13}
\gamma m_{0}c^{2}=\gamma \hbar \omega _{0}
\end{equation}
and
\begin{equation}
\label{II.14}
p'=\gamma m_{0}c=\hbar \gamma \beta k_{0.}
\end{equation}
From Eq. (\ref{II.14}), \( \gamma \beta k_{0}, \) can be identified as the
DeBroglie wave vector from conventional QM. Note that in expression (\ref{II.12}),
the carrier propagates with the ``phase velocity'' \( v \) while the modulation
has the ``group velocity'' \( V_{group}=c^{2}/v \). In the usual QM interpretation,
on the other hand, \( v \) is identified as a group velocity for a localized
wave packet intended to represent the local character of a particle. In that
case the dispersion relation \( V_{group}=v=\partial \omega /\partial k, \)
when integrated, yields the equation \( E_{kinetic}=m_{0}v^{2}/2=\hbar \omega _{D} \),
which relates the particle's kinetic energy \( E \) to a frequency \( \omega _{D} \),
which is identified as the temporal frequency of the particle's DeBroglie wave.
A DeBroglie wave packet is not required, however, to account for the wave-like
behavior of a localized particle, if wave-like properties can just as well be
ascribed to modulated \emph{Zitterbewegung}. More importantly, this new interpretation
suggests direct resolutions for many of the semantic and epistemological problems
vexing other interpretations of QM as discussed below in Section IV.

\section{Time Evolution in Quantum Mechanics }

Time evolution in QM is governed by the Schr\"odinger Equation. Being a hyperbolic
differential equation, the Schr\"odinger Equation has solutions which are time
reversible, and in this respect, \emph{inter alia}, are structurally distinct
from those for the parabolic diffusion equation. This fact, which arises in
various and not always obviously related manifestations, precludes interpreting
that part of QM connected with the Schr\"odinger Equation in terms of Stochastic
Mechanics, where diffusion processes govern the time evolution of physical phenomena.\footnote{
This point is obfuscated by the fact that propagation formulae pertaining to
deterministic processes also can be cast to have the apparent structure of Fokker-Planck
equations describing the time evolution of stochastic processes. In these cases,
however, if fundamental assumptions reintroducing the epistemological problems
of QM in new forms are eschewed, the physical ``diffusion constant'' can be
shown on the basis of kinematics alone to equal zero. 
} The conceptual model developed above, \emph{viz.} of the DeBroglie wavelength
as a manifestation of kinematically derived modulation on \emph{Zitterbewegung}
driven by Lorentz invariant background radiation, however, supports an interpretation
for the Schr\"odinger Equation with a new and coherent intuitive physical motivation
based on classical physics. This new physical interpretation respects the fundamental
difference between effects attributable to \emph{Zitterbewegung} and described
by Second Quantization and effects described by basic QM or First Quantization,
including ``hyperbolic'' time evolution. Consider a particle subject to a force
\( \mathbf{F} \) and for which the density of trajectories on phase space is
\( \rho (\mathbf{x},\, \mathbf{p},\, t) \), where \( \rho (\mathbf{x},\, \mathbf{p},\, t=0) \)
can be regarded either as the distribution of initial conditions for similarly
prepared particles, or, equivalently, as the \emph{a priori} probability distribution
of initial conditions for a single particle. Time evolution of \( \rho (\mathbf{x},\, \mathbf{p},\, t) \),
is governed by the Liouville Equation
\begin{equation}
\label{III.1}
\frac{\partial \rho }{\partial t}=-\nabla \rho \cdot \frac{\mathbf{p}}{m}+(\nabla _{p}\rho )\cdot \mathbf{F},\; \; \; [\nabla _{p}\equiv \sum _{i=x,y,z}\frac{\partial }{\partial p_{i}}]
\end{equation}
By virtue of Eq. (\ref{II.14}), each value of \( p \) in \( \rho (\mathbf{x},\, \mathbf{p},\, t) \)
is correlated with a particular wavelength of kinematical modulation. As proposed
above, boundaries and geometrical constraints on the waves to which the particle
is tuned cause these waves to diffract and interfere so that gradients in their
energy densities are induced. These gradients, in turn, result in spatial variations
in the magnitude of the \emph{Zitterbewegung,} the averaged effect of which
is to systematically modify particle trajectories. Because the energy density
of a wave is proportional to the square of its intensity, the wavelength of
energy density oscillations caused by the modulation will be half that of the
modulation itself. That is, the wavelength of the physical agent modifying trajectories,
an energy gradient, is half that of the modulation. Further, an ensemble consisting
of multiple particles, either conceptual or extant, will be guided by an ensemble
of energy density waves derived from an ensemble of kinematical modulations.
The spatial structure of this ensemble wave is found by taking the Fourier transform
of \( \rho (\mathbf{x},\, \mathbf{p},\, t) \) with respect to \( 2\, \mathbf{p}/\hbar  \),
the wave vector for the physical agent; i.e.:

\begin{equation}
\label{III.2}
\widehat{\rho }(\mathbf{x},\, \mathbf{x}',\, t)=\int e^{\frac{i\, 2\, \mathbf{p}\cdot \mathbf{x}'}{\hbar }}\rho (\mathbf{x},\, \mathbf{p},\, t)d\mathbf{p},
\end{equation}
for which the similarly transformed Liouville Eq. is
\begin{equation}
\label{III.3}
\frac{\partial \widehat{\rho }}{\partial t}=\left( \frac{\hbar }{i\, 2\, m}\right) \nabla '\nabla \widehat{\rho }-\left( \frac{i\, 2}{\hbar }\right) (\mathbf{x}'\cdot \mathbf{F})\widehat{\rho }.
\end{equation}

Solutions for equations of this form are sought by first separating variables
using a transformation of the form
\begin{equation}
\label{III.4}
\mathbf{r}=\mathbf{x}+\mathbf{x}',\; \; \; \mathbf{r}'=\mathbf{x}-\mathbf{x}',
\end{equation}
which yields
\begin{equation}
\label{III.5}
\frac{\partial \widehat{\rho }}{\partial t}=\left( \frac{\hbar }{i\, 2\, m}\right) (\nabla ^{2}-(\nabla ')^{2})\widehat{\rho }-\left( \frac{i}{\hbar }\right) (\mathbf{r}-\mathbf{r}')\cdot \mathbf{F}\left( \frac{\mathbf{r}+\mathbf{r}'}{2}\right) \widehat{\rho }.
\end{equation}

In general, for an arbitrary form of the force  \( \mathbf{F}(\mathbf{r},\, \mathbf{r}') \),
this equation still is not separable. However, for the purpose of calculating
expectation values, solutions to Eq. (\ref{III.5}) are needed only on the line
in the \( \mathbf{r},\, \mathbf{r}' \) plane for which \( \mathbf{r}=\mathbf{r}' \).
For example, the expectation of \( \mathbf{x} \) is 
\begin{equation}
\label{III.6}
<\mathbf{x}>=\int \int \mathbf{x}\, \rho (\mathbf{x},\, \mathbf{p})dxdp,
\end{equation}
which, with the inverse Fourier transform of Eq.(\ref{III.2}), becomes
\begin{equation}
\label{III.7}
<\mathbf{x}>=\int \int \int \mathbf{e}^{\frac{-i\, 2\, \mathbf{p}\cdot \mathbf{x}'}{\hbar }}\mathbf{x}\, \widehat{\rho }(\mathbf{x},\, \mathbf{x}')dxdpdx',
\end{equation}
so that the transformations of variables, Eqs. (\ref{III.4}), give
\begin{equation}
\label{III.8}
<\mathbf{x}>=\int \int \delta (\mathbf{r}-\mathbf{r}')\left( \frac{\mathbf{r}+\mathbf{r}'}{2}\right) \widehat{\rho }(\mathbf{r},\, \mathbf{r}')drdr'
\end{equation}
where \( \delta (\mathbf{r}-\mathbf{r}') \) is a Dirac delta function that
restricts the solutions of Eq. (\ref{III.5}) to contributing only along the
line \( \mathbf{r}=\mathbf{r}' \) to the integral for calculating \( <\mathbf{x}> \).
Similar results are obtained fo  \( <\mathbf{p}> \)and functions of \textbf{x}
and \( \mathbf{p} \); e.g., the expectation of the energy, \( <\mathbf{p}^{2}>/2m. \)\footnote{
The manipulations used here to compute expectation values can be shown to lead
naturally to unique, symmetric ``operator'' expressions, thereby obviating \emph{ad
hoc} ``Hermitezation'' rules for the primitive operator equivalents.\cite{6} 
} Thus, techniques yielding separable solutions to Eq. (\ref{III.5}) which coincide
with solutions to the nonseparable equation on this line, may be used for this
limited purpose. 

These special solutions on the line \( \mathbf{r}=\mathbf{r}' \) can be found
by exploiting the symmetry of Eq. (\ref{III.5} in the variables \( \mathbf{r} \)
and \( \mathbf{r}' \) and the fact that \( \widehat{\rho }(\mathbf{r},\, \mathbf{r}') \)
is real to write it the form \( \psi ^{*}(\mathbf{r}')\psi (\mathbf{r}). \)Thus,
with the usual manipulations to separate variables, Eq. (\ref{III.5}) yields
\begin{equation}
\label{III.9}
\frac{\partial \psi }{\partial t}-\left( \frac{\hbar }{i\, 2\, m}\right) \nabla ^{2}\psi (\mathbf{r})+\left( \frac{i}{\hbar }\right) \mathbf{r}\cdot \mathbf{F}\left( \frac{\mathbf{r}+\mathbf{r}'}{2}\right) \psi (\mathbf{r})=\left( \frac{i}{\hbar }\right) \cal {F(\mathbf{r},\, \mathbf{r}')\psi (\mathbf{r})}
\end{equation}
and its complex conjugate, where \( i\cal {F}(\mathbf{r},\, \mathbf{r}')/\mathit {\hbar } \),
which plays the role of what could be called a ``function of separation,'' has
a general form determined by symmetry and dimensionality considerations. 

If \( \psi (\mathbf{x}) \) satisfies Eq. (\ref{III.9}) when \( \mathbf{r}\to \mathbf{r}' \),
then \( \psi ^{*}(\mathbf{r}')\psi (\mathbf{r}) \) will satisfy
\begin{equation}
\label{III.10}
\left\{ \frac{\partial }{\partial t}-\left( \frac{\hbar }{i\, 2\, m}\right)
(\nabla ^{2}-(\nabla ')^{2})-\left( \frac{i}{\hbar }\right)
(\mathbf{r}-\mathbf{r}')\cdot \mathbf{F}\left(
\frac{\mathbf{r}+\mathbf{r}'}{2}\right) \right\} \widehat{\rho }\equiv \{\cal
{O}\}\widehat{\rho }=\mathrm{0}
\end{equation}
on the line \( \mathbf{r}=\mathbf{r}' \). This is seen by writing \( \rho (\mathbf{r},\, \mathbf{r}') \)
as \( \psi ^{*}(\mathbf{r}')\psi (\mathbf{r}) \) in Eq. (\ref{III.10}), adding
zero in the form \( (\cal {F}-\cal {F})\psi ^{*}(\mathbf{r}')\psi (\mathbf{r}) \)
and rearranging to obtain 
\begin{equation}
\label{III.11}
\psi ^{*}\left\{ \cal {O}-\mathit {\left( \frac{i}{\hbar }\right) }\cal {F}\right\} \psi +\psi \left\{ \cal {O}^{*}+\mathit {\left( \frac{i}{\hbar }\right) }\cal {F}\right\} \psi ^{*},
\end{equation}
from which the conclusion follows by letting \( \mathbf{r}\to \mathbf{r}' \).
The specific form of \( \cal {F}(\mathbf{r}) \) necessary to preserve the physical
content of the Liouville can be determined by requiring
\begin{equation}
\label{III.12}
\frac{d<\mathbf{p}>}{dt}=<\mathbf{F}>
\end{equation}
to hold. This stipulation (which, as physics, adds nothing new but as mathematics
extracts structure obscured by the separation technique) leads to
\begin{equation}
\label{III.13}
\int \psi ^{*}(\mathbf{r}\cdot \nabla \mathbf{F}-\nabla \cal {F})\psi \,
\mathit{d}\mathbf{r}=\mathrm{0}
\end{equation}
which, with the vector identity \( \mathbf{r}\cdot \nabla \mathbf{F}-\nabla \cal {F}=\nabla (\mathbf{r}\cdot \mathbf{F}-\cal {F})-\mathbf{F} \)
implies that \( \cal {F} \) must satisfy
\begin{equation}
\label{III.14}
\mathbf{r}\cdot \mathbf{F}-\cal {F}=-\mathrm{V},\; \; \; [\mathbf{F}\equiv -\nabla \mathrm{V}].
\end{equation}
Finally, with this result, on the line \( \mathbf{r}=\mathbf{r}' \), Eq. (\ref{III.9})
becomes the Schr\"odinger Equation
\begin{equation}
\label{III.15}
i\hbar \frac{\partial \psi }{\partial t}=\frac{\hbar ^{2}}{2\, m}\nabla ^{2}\psi +V\psi .
\end{equation}

Consistency with the assumptions leading to the extraction\footnote{
The word ``extraction'' is used here deliberately in the belief that expository
physics should eschew mathematical terminology (e. g., \noun{Theorem, Proof,}
even \noun{Derivation}\emph{\noun{,}} \emph{etc}\noun{.}\emph{\noun{)}}
as it invites the inference of a degree of logical rigor, which in the face
of the vast complexity of nature is often unobtainable or yields results idealized
into irrelevancy. Moreover, it is probably wise not to read physical significance
into these or similar symbolic manipulations that has not been invested deliberately
and explicitly; indeed, conceiving mathematical models for natural phenomena
may be less an exploration of Nature (the privilege of the experimentalist)
than an exploration of one's own imagination.
} of the Schr\"o\-ding\-er Equation above, requires that its interpretable solutions
be those for which the resulting phase space density is everywhere positive
and such that \( \rho (\mathbf{x},\, \mathbf{p},\, t=0) \) is the appropriate
initial condition. The relationship between solutions satisfying these physics
requirements and the eigenfunctions of the Schr\"odinger Equation is a complex
matter. Although crucial to an ultimate judgement of the coherence of the interpretation
espoused herein, it is left for future study. However, it seems highly portent
in this regard, that ``thermal states;'' i.e., mixed states with Boltzmann weighting
factors, and coherent states, give physically interpretable, everywhere positive
densities among other desirable traits.\cite{7}

For potentials having the form of a quadratic polynomial, the extraction of
the Schr\"odinger Equation given above requires no unusual separation technique.
Since such potential include that the harmonic oscillator, a substantial fraction
of all analytically soluble problems are covered; but, unfortunately the most
important case of all, the Coulomb potential, is not included. The manipulations
used above to obtain the Schr\"odinger Equation are essentially the reverse
of those suggested long ago in a study of QM phase space densities.\cite{8}
The principle difference with the reasoning herein is that formerly it was assumed
that QM is the fundamental theory and that the solutions of the Schr\"odinger
Equation all represent realizable states although some yield nonsensical phase
space densities. (Also, nonquadratic potentials complicate correspondence with
the Liouville Equation.) Here, the Liouville Equation is given logical primacy
and only those solutions of the Schr\"odinger Equation are admitted which yield
meaningful Liouville densities.

In summary, the considerations above suggest that at the level of ``First Quantization,''
QM can be given a formal or mathematical structure in terms of what might be
called a theory of `Eikonal Mechanics.'\footnote{
`Wave Mechanics' would be as good a name were it not already virtually synonymous
with all of QM, including Second Quantization. The use here of either term is
meant to emphasize the fact that although the Schr\"odinger Equation contains
essentially the same dynamical laws as the Liouville Equation, it is formulated
especially to account for the ray-like nature of particle trajectories induced
by relativistic, kinematical modulation. 
}The dynamical laws embodied in Eikonal Mechanics are deterministic; statistical
aspects of the theory are conceptual in nature and pertain to probabilities
implicit in a density on phase space. At its foundation, therefore, Eikonal
Mechanics is incomplete because probabilities are an artifact of the theory,
not of nature. Underlying Eikonal Mechanics are the stochastic effects of \emph{Zitterbewegung};
at this deeper level probabilities are conceptually the same as those in the
study of Brownian motion in statistical physics, so that here also the theory
is incomplete. Thus, QM in this rendition is interpreted as a statistical theory
with built-in structure to account for alterations to the classical motion of
particles that endows their averaged trajectories with ray-like properties.
These alterations are a particular manifestation of the influence of all other
charged particles in the universe via the mediation of a residue of electromagnetic
background radiation.

\section{Comments on Interpretational Paradoxes}

Kinematical modulation as embodied in Eikonal Mechanics provides a physical
model for the phenomena described by First Quantization which offers new perspectives
on philosophical problems intrinsic to most widely considered interpretations
of QM. Many of the seemingly novel features of this new interpretation already
have been found heuristically to be implicit in the structure of QM if the same
degree of logical consistency is enforced for the semantics of the terminology
applied to its symbols as is required mathematically of the relationships among
the symbols themselves. These features, therefore, are explicit elements in
certain nonstandard interpretations, and \emph{ipso facto}, implicit in the
most widely endorsed, ``Copenhagen,'' interpretation. 

Although much has been said concerning paradoxes and ambiguities in the orthodox
interpretation of QM and the attendant measurement process, perhaps one of Einstein's
Gedanken experiments still captures the kernel of the problem in the most concise
and revealing fashion.\cite{9} Following Einstein, consider a beam of particles
impinging on a virtually infinitesimal hole in a barrier. According to the principles
of QM, upon passage through the hole the (plane wave) beam diffracts to become
a semispherical wave. Now, if a semispherical detector, centered on the hole,
were used to detect the diffracted beam, in time the observed result would be
a uniform distribution of distinct, point-impacts over the surface of the detector.
If, as Einstein observed, a QM wave function constitutes the ontological essence
of individual particles, then, although immediately preceding each impact the
wave function for each particle is finite over the whole surface of the detector,
at the instant of the impact it must ``collapse'' to a delta function at the
impact point. A collapse or focusing of wave functions, occurring even instantaneously,
is unobjectionable if wave functions are only mathematical artifacts for calculating
expectation values and not the embodiment of the material essence of particles.
However, purely formal, nonphysical or ``mathematical'' interpretations of wave
functions seem to be precluded by the fact that QM wave functions, like electromagnetic
radiation and waves in physical media, interact with material boundaries. In
Einstein's Gedanken experiment, for example, the wave function diffracts upon
passage through the hole, implying that it has substantive existence. 

Eikonal Mechanics resolves this conflict by identifying wave functions as indeed
just mathematical expressions for computing expectation values on phase space;
but, with the difference that, these expectation values pertain to special trajectories,
namely those which are ray-like in response to spatial patterns in the energy
density of kinematically modulated background waves. The physical work of diffracting
the trajectories of localized material particles is accomplished by the external
agent of modulated \emph{Zitterbewegung} acting on them like a guiding hand.
The whole effect is embedded in the mathematical structure of QM, specifically
the Schr\"odinger Equation, for which the solutions give phase space densities
of the physically feasible set of ray-like trajectories available to particles
or systems. Thus, measurement is not an ontological issue for Eikonal Mechanics
because it is only a ``recalibration'' of phase space densities, essentially
revising the initial conditions. Measurement has the same role here as is has
in classical statistical physics, where it only reduces the degree of ignorance
but does not purport to revamp realities instantaneously. 

The imagery of Eikonal Mechanics has a certain resemblance to propositions independently
derived in other studies which strive to render the interpretation of QM semantically
consistent. In particular, the concept of modulation waves was anticipated by
the DeBroglie-Bohm theory of the ``pilot wave.''\cite{10}\cite{11} The conceptual
difference between the pilot wave theory and Eikonal Mechanics is that whereas
the pilot wave is envisioned as part of the physical essence of a particle---emanating
from it in the manner of an ethereal scout, or conversely as the carrier of
the particle as a confined high density region---modulation waves are considered
to be in existence in the particle's environment \emph{a priori} and emanate,
so to speak, from all other particles in the universe. 

Another aspect of modulation waves was anticipated in the ``many worlds'' interpretation
of QM.\cite{12} In this interpretation wave function collapse is obviated by
positing\footnote{
Actually, EGW deduced the many-worlds interpretation is mandated by logic if
QM as formulated is assumed to be fundamental and complete, because then, in
principle, there must be a wave function for the whole universe for which no
external agent exists to cause collapse. 
}that each possible trajectory for the universe (and individual particles in
particular) exists such that at each juncture where there are diverse potential
outcomes, the universe (including every particle individually) replicates with
one ``daughter'' fulfilling each potentiality.\footnote{
Specifically, EWG analysis originally identified each measurement act to be
a juncture at which replication occurred; but, presumably only an insignificant
minority of such junctures actually could involve conscience observers. 
}The interpretation based on Eikonal Mechanics suggests that while the geometry
of a particular situation accommodates the existence of a bundle of possible
ray-like orbits available to particulate systems, individual systems in effect
are bumped stochastically by background radiation from orbit to orbit while
maintaining their individual identity. In other words, the multiplicity implicit
in QM that was found by EWG is located in the potentiality of the background
and not the actuality of material particles. 

In the contemporary literature on the interpretation of QM, the focus is often
on Bell's Theorem which has been credited with showing that there is 

\begin{quote} ``\ldots{}a contradiction between several principles we had wished
could all be true, including the validity of quantum theory, Einstein's relativity
principle, causality acting forward in time, local objective reality and the
structure of space{[}; but,{]} experiments support some of these principles
to a certain degree of accuracy only. The contradiction can be resolved by abandoning
one or several of these principles.''\cite{13} \end{quote}

In view of this statement, the question immediately arrises: within the model
based on Eikonal Mechanics, which of these principles has been abandoned? 

The answer provided by Eikonal Mechanics is: none at a fundamental level, although
the mathematical or formal structure of QM abandons local objective reality;
i.e., the local identity of particles in favor of extended densities conditioned
instantaneously by distant boundaries. This can be understood as follows. Consider
a particle in a box. According to Eikonal Mechanics, a particle is a localized
entity which is in equilibrium with background waves standing in the box. This
means that the behavior of a particle at a particular space-time point in a
box is affected by waves whose structure is determined, in part, by boundary
conditions; i.e., the sides of the box, located at space-like separations from
the particle. Here is the source of the nonlocal aspect of QM which seems to
violate local reality or time-like causality. In the imagery of Eikonal Mechanics,
however, the standing waves in the box result from background radiation emitted
in the past by other sources at null or light-like separations. The nonlocal
feature in QM, according to this interpretation, is the same as that in any
steady state solution to wave equations where boundary conditions determine
the solution in a region of interest. Therefore, according to this interpretation,
although particle densities on phase space do manifest nonlocal effects, such
effects do not constitute an ontological problem because the radiation inducing
these effects propagated along light-like separations in Minkowski Space from
the past. Phenomena peculiar to QM can be considered to be ``multibody effects''
caused by classical interactions from all other charged particles in the universe
rather than manifestations of the fundamental nature of particle interactions
exclusively. 

Underlying the relatively gross behavior of particles as described by Eikonal
Mechanics is the fine-grained \emph{Zitterbewegung}. All phenomena described
by Eikonal Mechanics, that is ``First Quatization,'' invite deeper analysis
to understand the ``anomalies'' and ``shifts'' caused by \emph{Zitterbewegung}.
And, at least in concept, this can be accomplished using SED for which there
is no problem with interpretation.

\section{Spin and External Fields}

To include external electromagnetic fields in this formalism, let the total
energy of the radiation to which the particle is exposed include that of the
background plus the potential energy from additional fields; i.e., let
\begin{equation}
\label{V.1}
m_{0}c^{2}=\hbar \omega +e\phi 
\end{equation}
where \( e \) is the charge on the particle and \( \phi (\mathbf{x}) \) is
the potential function for the additional field. The gauge for the external
field has been chosen so that the four-vector potential in the rest frame of
the particle is purely scalar or time-like. This is tantamount to assuming that
the potential energy of the additional fields contribute to the rest mass of
the particle such that \( (m_{0})_{total}=m_{0}+e\phi /c^{2} \).\cite{14} 

While clear as formalism, these considerations depend on the underlying theory
of electromagnetic interaction, in particular the use of potential functions
to take full relativistic account of interaction. Reference 13 provides reason
to suspect that there is considerable room for deeper understanding of the use
of potential functions consistent with special relativity. Nevertheless, the
role of Eq. (\ref{V.1}) as an axiomatic proposition preceding the logic and
manipulations used to extract the Schr\"odinger Equation, elucidates the structure
of QM by indicating that obscurities in the physical meaning of electromagnetic
potentials in QM are not wholly intrinsic to QM \emph{per se,} but substantially
to relativistic kinematics and dynamics. 

Intuitive motivation for spin consistent with concepts of Eikonal Mechanics
can be found as a manifestation of the existence of two independent polarization
states for electromagnetic radiation. This can be seen by conceptually decomposing
each background mode into its two polarization states and considering that a
given density on phase space \( \rho (\mathbf{x},\, \mathbf{p},\, t) \) consists
of two parts, \( \rho _{1} \) and \( \rho _{2} \), each pertaining to that
portion of an ensemble in equilibrium with one polarization state. Time evolution
of a tandem set of densities is governed by a tandem form of the Liouville Equation
\begin{equation}
\label{V.2}
\overrightarrow{1}\left[ \begin{array}{c}
\partial \rho _{1}/\partial t\\
\partial \rho _{2}/\partial t
\end{array}\right] =\left[ (\sigma \cdot \frac{d\mathbf{x}}{dt})(\sigma \cdot \nabla )+(\sigma \cdot \mathbf{F})\right] \left[ \begin{array}{c}
\rho _{1}\\
\rho _{2}
\end{array}\right] 
\end{equation}
where \( \sigma  \) and \( \overrightarrow{1} \) are the Pauli spin matrices
and the \( 2\times 2 \) identity matrix respectively. By using the 3-vector
identity \( (\sigma \cdot \mathbf{V})((\sigma \cdot \mathbf{W})=\mathbf{V}\cdot \mathbf{W}+i\sigma \cdot (\mathbf{V}\times \mathbf{W}) \)
together with the equation for energy equilibrium in the presence of additional
fields, Eq. (\ref{V.1}), and the methods of Fourier decomposition of modulation
waves as presented in Section III above, straightforward but somewhat tedious
manipulations yield the Pauli version of the Schr\"odinger Equation for Fermions
\begin{equation}
\label{V.3}
i\hbar \frac{\partial \psi }{\partial t}=\frac{1}{2m}\left( \sigma \cdot (i\hbar \nabla -\frac{e\mathbf{A}}{\mathbf{c}})\right) ^{2}\psi +\phi \psi ,
\end{equation}
where \( \mathbf{A} \) is the electromagnetic vector potential associated with
\( \phi  \) Eq. (\ref{V.1}). 

Except for anomalies to be taken into account by Second Quantization or SED
and caused by \emph{Zitterbewegung,} as is well known, Eq. (\ref{V.3}) gives
the correct magnetic moment for the electron. (In spite of prejudicial nomenclature,
herein spin matrices are actually just the vehicles for attaching the algebraic
structure implicit in what, in the language of Differential Geometry, is a two
dimensional bundle attached to each independent tangent fiber at every point
of a three dimensional manifold. A two-dimensional structure is needed to take
account of the two polarization states.) Bosons, as usual, are comprised of
multiple Fermions. The imagery suggested by these manipulations implies that
spin magnetic moments result from currents attendant to \emph{Zitterbewegung}
motion. In the absence of magnetic fields, particles equilibrate randomly with
both polarization modes in the background. An external magnetic field, however,
will lift the degeneracy by causing those particles whose instantaneous random
motion is, when projected on the oriented plane perpendicular to the axis of
the magnetic field, positive rotation, to precess so as to tend to couple preferentially
to, and thereafter maintain energetic equilibrium with a particular circular
polarization state of that background mode propagating parallel to the axis
of the magnetic field. Those particles with the opposite sense of rotation will
likewise couple with the opposite polarization state. Thus, the binary result
of Stern Gerlach experiments with respect to an arbitrary axis can be understood
as a consequence of the fact that the magnetic field itself is responsible for
inducing the axis of precession with respect to which the background mode is
decomposable as the sum of two circular polarization states, each of which is
then independently available for energetic equilibrium.

\section{Bell's Theorem }

In analysis of the foundations of QM, Bell's Theorem has taken a central position
because it provides an experimentally testable distinction between QM and possible
alternate formulations involving hidden variables. Thus, it is essential to
analyse its significance from the vantage of Eikonal Mechanics. 

Recall that the proof of Bell's Theorem is formulated in terms of the disintegration
of a spin free boson into two fermions of opposite spin. This event is particularly
significant because it is also a realizable demonstration of a situation originally
analyzed by Einstein, Padolski and Rosen (EPR) to demonstrate the incomplete
character of QM. They noted, in effect, that according to conventional QM, prior
to observation, the wave function for each derivative particle or ``daughter''
comprises an ambiguous mixed state which is collapsed upon measurement to a
specific, observed value. Thus, when one daughter is measured, by virtue of
spin conservation, the wave function for her sibling is fixed instantly. If
QM is complete and fundamental, then the objective state of each daughter must
also be ambiguous prior to a measurement of either daughter and this, in turn,
implies that the objective states of both daughters, at space-like separation
from each other, are altered; i.e., rendered specific, by a spin measurement
made on only one. 

EPR then argued, as is well known, that QM must be incomplete because, by admitting
ambiguous mixed states that require instantaneous transmittal of information
from one particle to its twin when a measurement is made on only one, it violates
Special Relativity. In concept, the problem with the EPR event is exactly the
same as that illustrated by a particle beam diffracting through an orifice as
discussed above in Section~IV. 

The proof of Bell's Theorem consists in showing that QM expectation values for
the correlation of spin measurements, \( P \), made with respect to arbitrary
axis denoted by \( \mathbf{a},\, \mathbf{b} \) and \( \mathbf{c} \) in the
realization of the decay version of the EPR event can not satisfy the inequality
\begin{equation}
\label{VI.1}
1+P(\mathbf{b},\, \mathbf{c})\geq |P(\mathbf{a},\, \mathbf{b})-P(\mathbf{a},\, \mathbf{c})|,
\end{equation}
which, as Bell found, should obtain for all such correlations of local, dichotomic
variables.\cite{1} The connection to the physics of the EPR event in the proof
of this Theorem is established by constraints imposed on the explicit form of
the correlations \( P \), to wit:
\begin{equation}
\label{VI.2}
P(\mathbf{a},\, \mathbf{b})\equiv \int d\lambda \, \rho (\lambda )A(\mathbf{a},\, \lambda )B(\mathbf{b},\, \lambda ),
\end{equation}
where \( \lambda  \) represents possible hidden variables, \( \rho (\lambda ) \)
a probability distribution over the variables \( \lambda  \), and \( A \)
and \( B \) are the results of measurements of the spin of the daughters. Now,
to duplicate QM results, \( A \) and \( B \) must be dichotomic; i.e., \( A=\pm 1,\; \; B=\pm 1 \).
As the measuring instruments; e.g., Stern-Gerlach magnets, are capable of registering
a continuum of values, this constraint fixes \( A \) and \( B \) as expressions
of the properties of the derivative particles. To be ``local'' requires that
\( A \) does not depend on \( \mathbf{b} \) and \( B \) does not depend on
\( \mathbf{a} \). Given these constraints, Eq. (\ref{VI.1}) follows directly.
The fact that the QM expectations calculated for the EPR event do not satisfy
Eq. (\ref{VI.1}) is taken to mean that theories with hidden variables can not
duplicate the structure of QM. Furthermore, experiments seem to verify QM rather
than Eq. (\ref{VI.1}). 

However, the relevance of the assumption that a spin measurement of either daughter
is independent of the other can be questioned. This assumption is reasonable
if it is assumed that each daughter departs the scene of the disintegration
in an ambiguous state so that either dichotomic outcome is possible before the
act of measurement itself breaks the ambiguity and precipitates a distinct value.
If, on the other hand, each daughter in fact departs the scene of the disintegration
with properties that unambiguously determine the value of the spin which will
be subsequently measured with respect to an arbitrary axis, then instantaneous
information transmission across space-like separations is not required to account
for the result. This is exactly the situation as envisioned from the vantage
of Eikonal Mechanics for which wave functions do not constitute the essence
of material existence, rather they are just symbolic expressions for computing
densities in the phase space of ray-like trajectories for what are distinct
material particles in the classical sense. 

Moreover, the proof of Bell's Theorem explicitly exploits the assumption that
spin correlations are to be computed with the dichotomic data resulting from
measurements. While seemingly reasonable, this appears not to be consistent
with the calculation carried out using QM rules. The QM result, 
\begin{equation}
\label{VI.3}
<\sigma _{1}\cdot \mathbf{a},\, \sigma _{2}\cdot \mathbf{b}>=-\mathbf{a}\cdot \mathbf{b},
\end{equation}
is obtained if \( A \) and \( B \) are taken as projections on the axes of
the measuring devices and dichotomicity imposed, so to speak, to interpret results
only \emph{after} (if at all) the correlation is computed. This may reflect
the fact that in orthodox QM, measurement is subordinated to an appendix of
the formalism.\footnote{
A recent publication, \cite{16}, criticized the interpretation of experiments
exploiting atomic cascade events to generate oppositely polarized photons to
test Bell's Theorem. The essence of the criticism is that properly accounting
for the spherical symmetry of photons so generated invalidates the conclusion
that such experiments support QM vice possible hidden variable theories. Even
without this criticism, however, a substantiation of Bell's result derived from
experiments on ``photons'' would be less satisfying that an experiment with
entities classically regarded as particles. Because of the existence of phenomena
such as wave packets, needles, and solitons for waves, particle-like behavior
by waves (especially interacting with \emph{particulate} detectors) is less
counterintuitive than wave-like behavior, in particular diffraction, by Gibbsian
ensembles of individual particles.
}

\section{Conclusion}

Besides their obvious utility for philosophical analysis and pedagogical aids,
interpretations for physics theories ideally serve to provide coherent mental
images and mutually consistent vocabulary in order to facilitate thinking about
and describing physical phenomena. Although they are not unequivocally required
to formulate or use operationally correct, quantitative physics theories---as
the history of QM has demonstrated abundantly---images and evocative vocabulary
nevertheless often provide guidance for modeling particular phenomena and inspiration
for new experimental, theoretical or calculation techniques. The acceptance
of a particular interpretation, therefore, is determined in part by the success
of specific models and calculations which it prompted. Nevertheless, more is
usually desired; a physical model should have universality and render all phenomena
within its presumed domain intuitively comprehensible. 

Perhaps the central challenge in this regard for an interpretation for QM based
on Eikonal Mechanics is to provide visualization or intuitive justification
for the Pauli exclusion principle, in particular as applied to atomic structure.
For a start, consider a pair of electrons in orbit about a nucleus where each
electron assumes a spin magnetic moment by virtue of exposure to the apparent
magnetic field generated by its orbital motion. The image of such an interacting
couplet seeking an energy minimum in antialignment, seems natural. Hence the
rule that electrons tend to form pairs with spin ``up and down'' for each set
of orbital quantum numbers. For increasing numbers of electrons, which repel
each other and shield the nucleus, the increase in total energy for the electron
collection and decrease of individual binding energies also seem intuitively
natural. 

The image of electrons segregated and stacked by energy eigenvalue as neatly
as implied by the exclusion principle, even if only into zones, is, however,
less natural. But, literal compliance with this principle may not be necessary.
All that is required to account for observed spectra, is that emission and absorption
occur as if electron orbits were on the average neatly ordered and singly populated
by eigenvalue. This could well occur in a macroscopic sample if the combined
effect of the background and boundary conditions is that the only radiation
to escape from the sample (or to fail to penetrate through it in the case of
absorption) is that which does not fit the equilibrium interplay between background
and atom. For electrons orbiting a point charge, spherical symmetry is the boundary
condition which determines the eigenfunction space and eigenvalues; i.e., it
structures the energy levels of nonequilibrium emission and adsorption. Thus,
in the steady state, individual electrons might be envisioned to be constantly
migrating through various eigenzones or exchanging orbits among themselves such
that on the average, energy is exchanged only with the background. Thus, in
this imagery, statistical characteristics are vested in the motion of individual
particles rather than in the population distribution among states each of which
is envisioned to comprise ensembles of fixed energy level systems locked into
specific orbits. This imagery, in fact, seems uniquely consistent with the previous
observation that physically interpretable densities seem to result only from
mixed states with all eigenfunctions present. Still, the question of just how
to determine specific weighting factors for eigenfunctions comprising physically
realizable mixed states is open but not obviously unresolvable. 

The problems with fundamental QM addressed herein are overwhelmingly verbal.
QM as a quantitative codification of physical phenomena has been verified virtually
beyond question. What is \emph{said} about reality based on QM, however, is
at least semantically inconsistent; one consequence of which is that QM defies
visualization. Eikonal Mechanics, as outlined above, proposes vocabulary with
intuitive images to portray reality in a new way on the basis of QM without
changes in its current mathematical (symbolic) formulation. New or reformulated
calculations simply to replace customary QM techniques, are in general not required;
the primary value of any new paradigm at this level is philosophical and pedagogical.
Hopefully, however, the intuitive concepts underlying Eikonal Mechanics, whether
or not they ultimately withstand critical analysis of their use to interpret
QM, can be exploited or extended in order to contribute to the explanation of
still ill understood physical phenomena or the discovery of utterly new ones.

Current address:\\
P. F. 2040\\
99401 Weimar, BRD\\
\texttt{kracklau@fossi.uni-weimar.de}
\end{document}